\documentclass[prl,amsmath,amssymb,dvips,twocolumn,showpacs]{revtex4} 
\usepackage{epsfig} 
\usepackage{bm} 
\usepackage{amsmath}
 
\newcommand{\be}{\begin{equation}} 
\newcommand{\ee}{\end{equation}} 
\newcommand{\bea}{\begin{eqnarray}} 
\newcommand{\eea}{\end{eqnarray}} 
\newcommand{\Tr}{\text{Tr}}

\def\half{\hbox{\rm${\frac{1}{2}}$}}

\begin{document} 
 
\title{Mixed-spin pairing condensates in heavy nuclei}
 
\author{Alexandros Gezerlis$^1$, G.F.~Bertsch$^{1,2}$, and Y.L.~Luo$^1$} 
\affiliation{$^1$Department of Physics, University of Washington, Seattle, Washington 98195--1560 USA} 
\affiliation{$^2$Institute for Nuclear Theory, University of Washington, Seattle, Washington 98195--1560 USA}

\begin{abstract} 

We show that the Bogoliubov-de Gennes equations for nuclear ground-state wave
functions support solutions in which the condensate has a 
mixture of spin-singlet and spin-triplet pairing.  We find that
such mixed-spin condensates do not occur when there are equal numbers
of neutrons and protons, but only when there is an isospin imbalance.
Using a phenomenological Hamiltonian, we predict
that such nuclei may occur in the physical region within the proton dripline. 
We also solve the Bogoliubov-de Gennes equations with variable
constraints on the spin-singlet and spin-triplet pairing amplitudes. For
nuclei that exhibit this new pairing behavior, the
resulting energy surface can be rather soft, suggesting that there may
be low-lying excitations associated with the spin mixing.

\end{abstract}

 
\maketitle

{\it Introduction.} The usual pairing found in nuclei is between identical nucleons in the
spin-singlet channel.  Although the spin-triplet interaction is stronger,
the spin-orbit field tends to suppress pairing in the triplet 
channel\cite{po98,sc10}.  However, spin-triplet pairing becomes favored
in nuclei with equal numbers of neutrons $N$ and protons $Z$ when the nucleon number
is very large (probably beyond the proton dripline)\cite{be10}.  In this 
work we address nuclei with $N\ne Z$ and find some
surprising results:  a)  the domain where spin-triplet pairing dominates
extends well off the $N=Z$ line; b) the condensate changes character smoothly
between pure spin-triplet on the $N=Z$ line to pure spin-singlet at large
neutron excess,  c) the mixed-spin nuclei that we find extend below the proton
dripline and are thus relevant to experiment.

{\it Context.}  The expectation that isospin-zero ($T=0$) neutron-proton
pairing should exist comes from the fact that the interaction in the
spin-triplet (isospin-singlet) channel, which binds the deuteron, is
stronger than the $^1S_0$ interaction that is largely responsible for
ordinary identical-particle spin-singlet pairing.  It was suggested a long
time ago that neutron-proton pairing is important near the $N=Z$ line (see
Refs. \cite{goo72,goo99} and works cited therein, as well as Refs.
\cite{ma00a,ma00b} for a discussion of the experimental situation).  A
number of theoretical works have examined the possibility that nuclei may
contain a $T=0$ spin-triplet neutron-proton (`deuteronlike') condensate
when $N=Z$, suggesting that states of high angular momentum might favor
$T=0$ pairing \cite{sa97,ter98,go01}. The possibility of mixed-spin
condensation, $T=0$ and $T=1$, has also been raised in Refs.
\cite{goo98,goo99} for $N=Z$ medium-mass nuclei, although no mixed-spin
ground states were shown.  In this Letter, we present our findings for
the existence of mixed-spin solutions to the Bogoliubov-de Gennes (BdG) equations
for the ground-state of large but accessible nuclei off the $N=Z$ line.

{\it Hamiltonian.} We use the same Hamiltonian here as was used
in Ref. \cite{be10}.  It contains a one-body and a two-body part represented
in Fock space as:
\be
\hat H = \sum_i \langle i| H_{sp}| j \rangle a^\dagger_i a_j +
\sum_{i>j,k>l}  \langle ij|v | kl \rangle a^\dagger_i a^\dagger_j a_l a_k
\ee
where $i,j$ label orbitals in a spherical shell-model basis.
The one-body part $H_{sp}$ is taken from the eigenstates of a Wood-Saxon
potential of standard form, containing a kinetic energy,
a potential well, and a spin-orbit term.  
The two-body interaction is of contact
form:
\be
\langle ij | v | kl \rangle = \hbox{\rm${\frac{1}{4}}$}\langle ij|(3v_t+v_s
+(v_t-v_s)\vec\sigma\cdot\vec\sigma')\delta^{(3)}(\vec r-\vec r')P_{L=0}
 | kl \rangle.
\label{eq:pot}
\ee
where $P_{L=0}$ projects onto the spherically symmetric part of the pair
wave function. This Hamiltonian is appropriate for
systems with no nuclear deformation, accenting the pairing condensates.
 There are two interaction strengths,
$v_t$ and $v_s$, corresponding to spin-triplet and spin-singlet,
respectively. These were determined by fitting
to phenomenological shell-model Hamiltonians.  The interaction of Eq. (\ref{eq:pot}) 
can generate
6 independent condensates, counting only spin and isospin quantum numbers.
We label these by an index $\alpha$ enumerated in Table I.
\begin{table}[htb]
\begin{center}
\begin{tabular}{|c|cccccc|}
\hline
$\alpha$ & 1& 2& 3& 4& 5& 6\\
\hline
$(S,S_z)$ &(0,0)&(0,0) &(0,0) &(1,1) &(1,0)&(1,-1)\\
$(T,T_z)$  &(1,1) &(1,0)&(1,-1)&(0,0)&(0,0) &(0,0)\\
\hline
\end{tabular}
\caption{Spin-isospin channels for pairing condensates.}
\end{center}
\end{table}

Finally, we note that the Coulomb interaction is omitted in the above
Hamiltonian.  The main effect of this is that the calculated nuclei are only
physical within the 
proton dripline.  Nevertheless, the pairing phenomena that can be elucidated
beyond the proton dripline are interesting on a purely theoretical level.  
Also, as we shall show,
the region where novel forms of pairing may occur extends into the physical region,
below the proton dripline.

{\it BdG theory.}  The Bogoliubov-de Gennes theory is defined
by  minimizing the Hamiltonian under Bogoliubov transformations of
the Fock-space vacuum, subject to constraints such as the neutron and
proton number expectation values.  In the notation of \cite{RS}, the 
Bogoliubov transformations are
parametrized by the matrices $U$ and $V$ giving the definition of
the quasiparticle operators in terms of the Fock-space
operators. The key equations
in the theory are the formulas for ordinary and anomalous densities,
$\rho = V^*V^t$ and  $\kappa = V^*U^t$, respectively,
and the formula for the expectation value of the Hamiltonian,
\be
\label{H00}
H^{00} = {\rm Tr}( \varepsilon \rho + \half \Gamma \rho - \hbox{\rm
${\frac{1}{2}}$} \Delta
\kappa^*).
\ee
As usual, the matrices $\Gamma,\Delta$ are defined through the 
standard relations
$
\Gamma_{ij} = \sum_{kl} v_{ikjl} \rho_{lk}$ and
$\Delta_{ij} = \half \sum_{kl} v_{ijkl} \kappa_{kl}$.
Here, and in Eq. (\ref{eq:grad}) below, superscripts denote the number of
quasiparticle creation and annihilation operators. Since our Hamiltonian
is phenomenological, we assume that the $\Gamma \rho$ term is included
in it. We therefore omit explicit consideration of it below.

{\it Calculational procedure.}  Traditionally the minimization is
carried out using the BdG equations, which are arrived at by
setting the variational derivative
of the energy with respect to $U$ and $V$ to zero. (Actually the
variation must be constrained to preserve the unitarity of the Bogoliubov
transformation.  This introduces Lagrange multipliers that give the BdG
equations their structure as eigenvalue equations for the quasiparticle
energies.)  The BdG equations are solved for some assumed
density, and the solution is used to update the density.  This process
is  iterated to self-consistency.  However, to study
the energetics with different types of condensates it is necessary to 
deal with many constraining fields and therefore thoroughly explore the space of
allowed Bogoliubov transformations.  Under these conditions, 
the BdG minimization is easier to carry out by the gradient method \cite{RS},
and we take advantage of that method here. 
In taking the variational derivative
of the Hamiltonian one makes use of the generalized Thouless 
matrix $Z$, whose elements are independent of each other.  The gradients 
of the Hamiltonian and the operators to be constrained can then be
applied to update a trial set of $U,V$ matrices, using the steepest descent or other
numerical methods\cite{ro11}.  The change in the expectation value of a one-body
operator $Q$ can be expressed:
\be
Q^{00}_{new} = Q^{00}_{old} - \Tr ( Q^{20} Z) + {\cal O}(Z^2).
\label{eq:grad}
\ee
A similar formula applies for the Hamiltonian,
since its expectation value can be expressed in terms of one-body
expectation values.  To insure that the space of possible
Bogoliubov transformations has been adequately explored, we carry out
the iteration process repeatedly starting from $U,V$ matrices obtained by
transformations from the vacuum or other states by $Z$ transformations. 
We have used the gradient method
to solve the BdG equations with 8 simultaneous constraining fields, 2 for the 
neutron and proton particle numbers, and 6 for the pairing amplitudes corresponding
to the 6 distinct channels of Table I. To be more precise, the 6 
constrained amplitudes are computed as $\kappa_\alpha=\Tr(P_\alpha \kappa)$ where the
matrices $P_\alpha$ are defined in terms of the quantum numbers
$(\ell_k,\ell_{zk},s_{zk},t_{zk})$ of the orbitals $k$ as
\be
 P_{\alpha,ij}  = \sqrt{2} 
\left(\half\, s_{zi}\, \half \,s_{zj} | S(\alpha) S_z(\alpha)\right)\times
\ee
$$
\left(\half\, t_{zi}\, \half\, t_{zj} | T(\alpha) T_z(\alpha)\right)
 (-)^{\ell_i-\ell_{zi}}
\delta_{\ell_i,\ell_j}\delta_{\ell_{zi},-\ell_{zj}}.
$$

In the computation, even-A and odd-A nuclei are distinguished by the number
parity of the Bogoluibov transformation\cite{ro11}.  For odd-A nuclei,
there is a block structure of the Hamiltonian and the odd number parity
is imposed on one of the blocks.  Each block must be tested to find the
global energy minimum.

{\it Results.}  A quantity that allows us to accurately gauge
the relative importance of the pairing condensates is the correlation
energy, $E_{corr} = E_0 - E$, where $E_0$ is the energy of the 
ground state in the absence of a pairing condensate, i.e. the result
of setting all $\kappa_\alpha$ to zero. 

\begin{figure}[t]
\vspace{0.5cm}
\begin{center}
\includegraphics[width=0.46\textwidth]{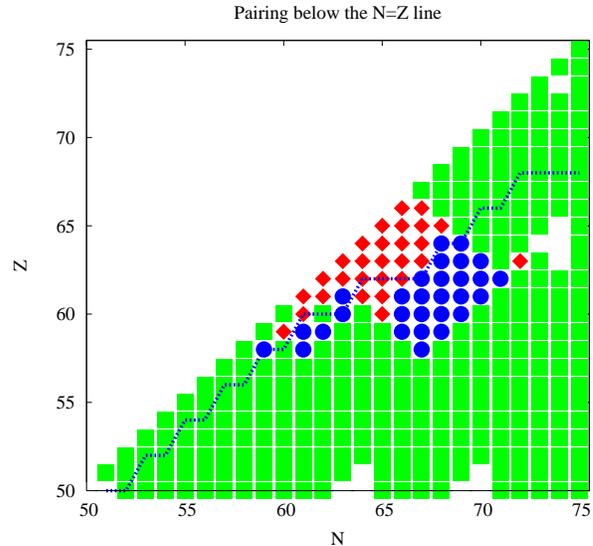}
\caption{(color online) 
Chart of nuclides with $Z \le N$ for neutron numbers from 50 to 75. Blank squares
denote nuclei that exhibit practically no pairing ($E_{corr} < 0.5$ MeV), green squares
signify the case where the pairing condensate is mostly spin-singlet, red
diamonds are used for the nuclei that exhibit spin-triplet pairing, while blue circles
denote nuclei for which the pairing is a mixture of spin-singlet and spin-triplet.
The blue dashed line is the proton-drip line from Ref. \cite{mo95}.}
\label{fig:chart}
\end{center}
\end{figure}

\begin{figure*}[t]
\centering
\begin{tabular}{ccc}
\epsfig{file=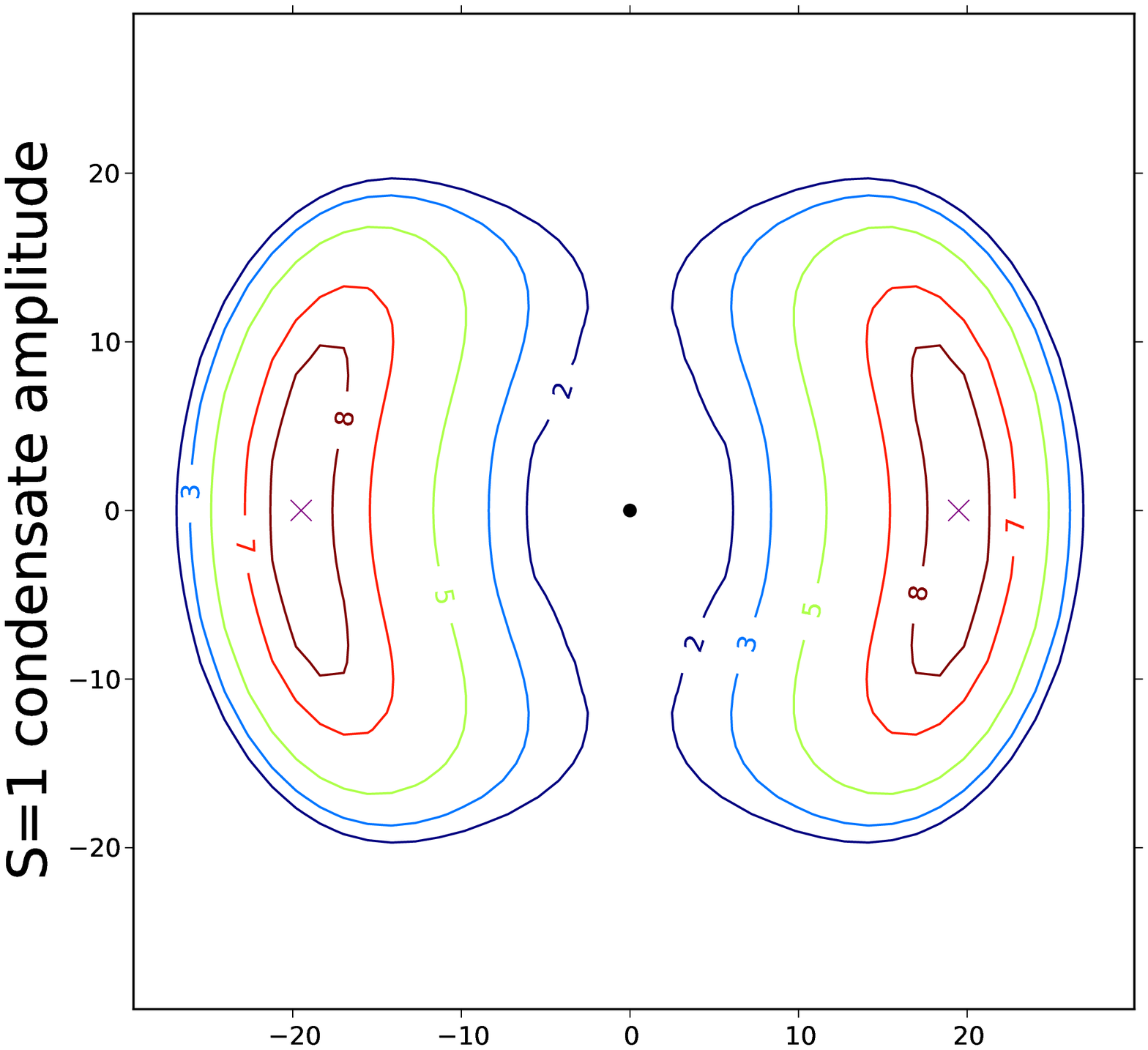,width=0.33\linewidth, viewport = 280 0 840 540,clip} &
\epsfig{file=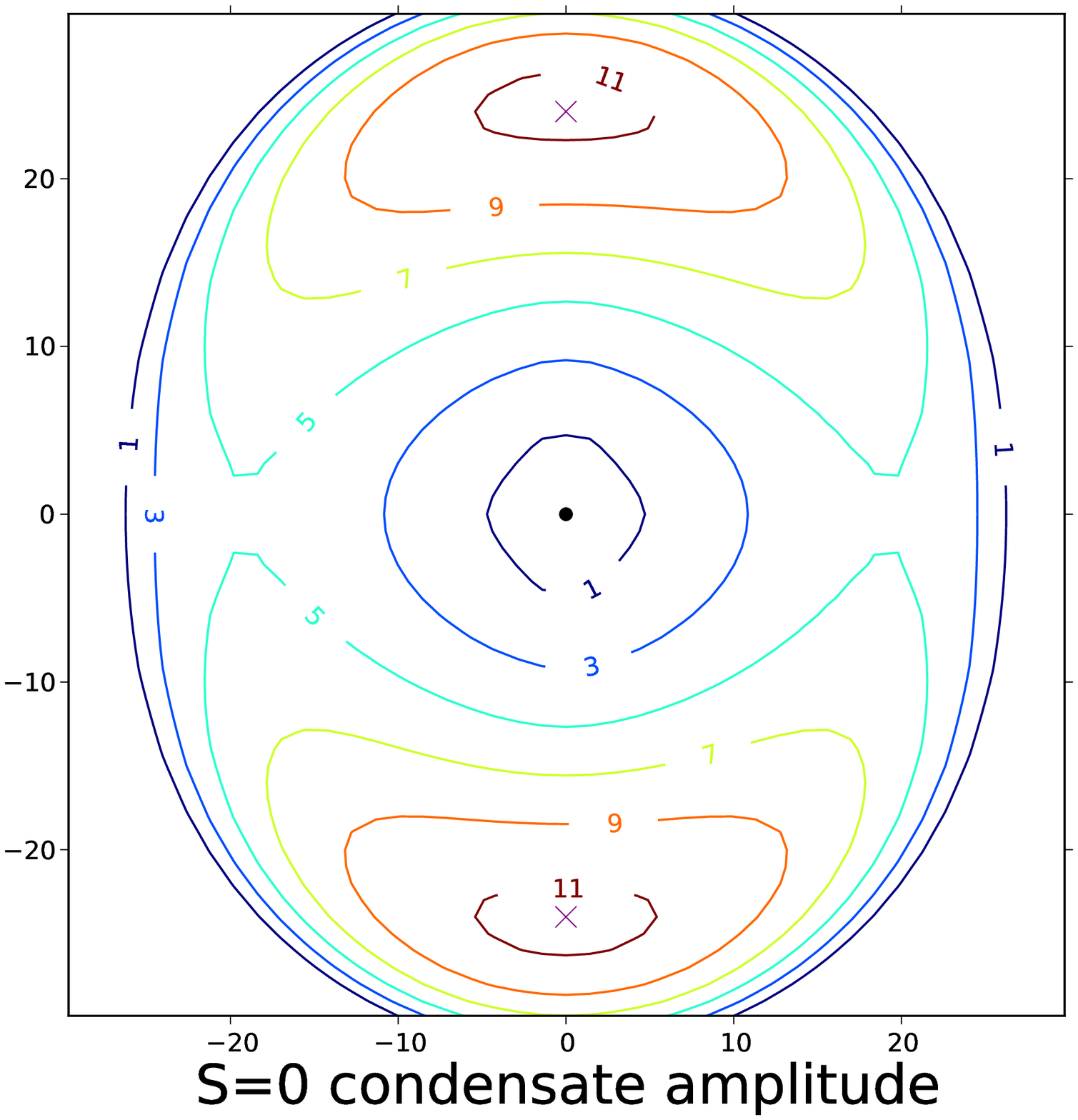,width=0.33\linewidth, viewport = 280 0 840 540,clip} &
\epsfig{file=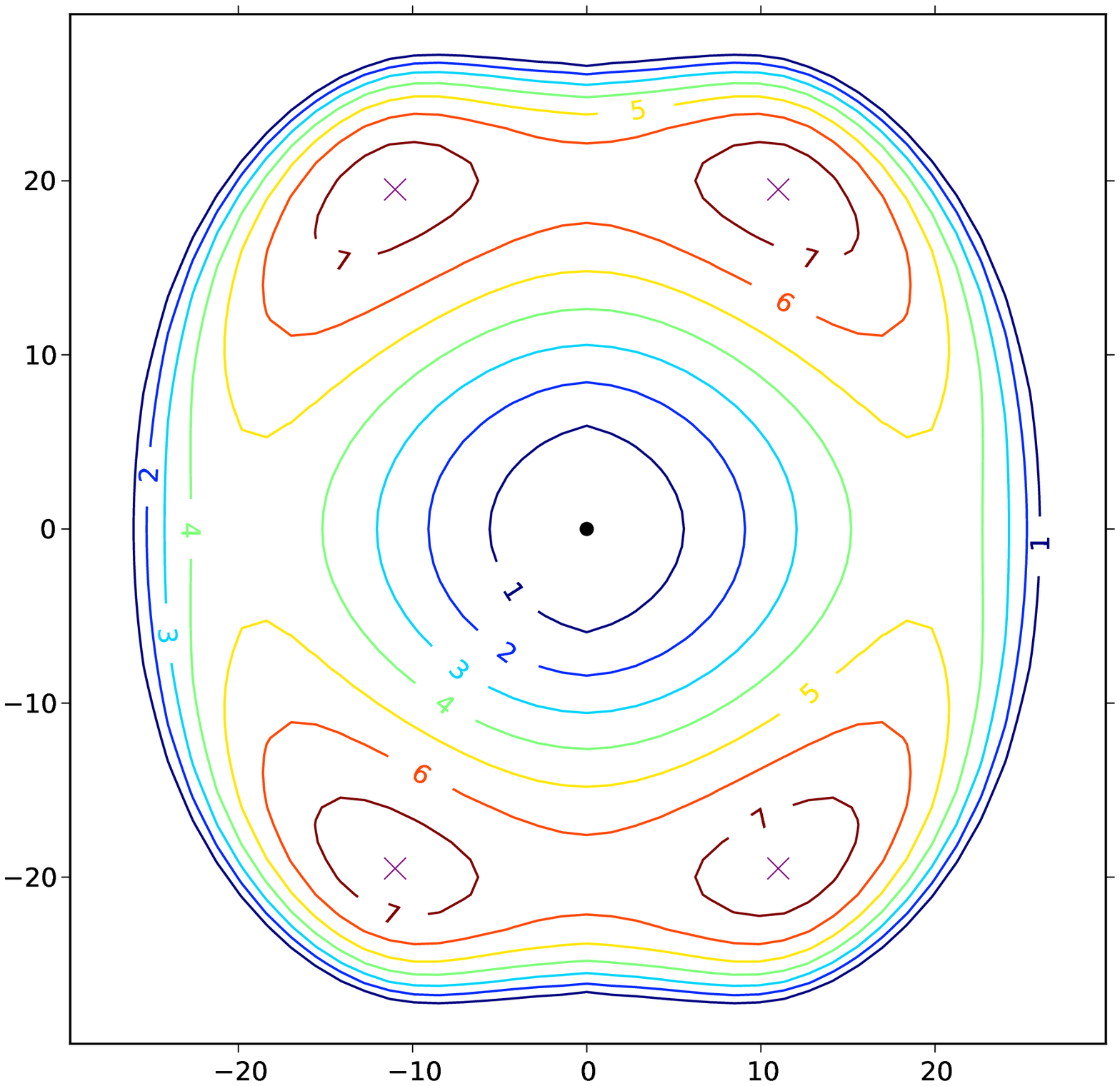,width=0.33\linewidth, viewport = 280 0 840 540,clip} \\
\end{tabular}
\caption{(color online) 
Contour plots of the correlation energy in three different $A = 132$ nuclei as a function
of the amplitudes of selected spin-singlet and spin-triplet condensates. 
Left panel: $^{132}_{60}$Nd, mainly spin-singlet pairing, as is common in most nuclei
studied to date; middle panel: $^{132}_{66}$Dy, mainly spin-triplet pairing, similarly to 
what was introduced in Ref. \cite{be10}; right panel: $^{132}_{64}$Gd, mixed-spin pairing
exemplifying a qualitatively new feature, namely the gradual crossover from spin-singlet 
to spin-triplet pairing. The numbers show correlation energies in MeV.
In all three cases, each peak is marked by an X.}
\label{fig:contour}
\end{figure*}

We have mapped out all nuclides with $Z\le N$ for neutron numbers from 50 to 75
and show the results in Fig. \ref{fig:chart}. A few nuclei have very small
correlation energies (white in Fig. \ref{fig:chart}), while the
majority of nuclei, above
and below the proton dripline, are spin-singlet (green squares in Fig.
\ref{fig:chart}). 
However, a group of nuclei with
neutron numbers roughly from 60 to 70 exhibit spin-triplet pairing
(red diamonds in Fig. \ref{fig:chart}), or as 
discussed below, a crossover between the two kinds of pairing.
The demarcation between the three kinds of pairing is somewhat arbitrary,
so we have chosen to call ``mixed-spin paired'' those nuclei for which
the spin-singlet amplitude is between one and three quarters of the total
pairing amplitude.
Note that the island of mixed-spin paired nuclei (blue circles in Fig. \ref{fig:chart})
contains many nuclei that lie within the proton dripline.  Thus, the
predicted mixed-spin pairing may be relevant to experimental investigation.  

In an attempt to understand how the pairing condensate changes
from spin-singlet to spin-triplet, we have examined in more detail
some nuclei at mass number $A=132$.  Here, the $N=Z$
nucleus  $^{132}_{66}$Dy  exhibits spin-triplet pairing.  Changing the
neutron-proton asymmetry, one reaches the region of ordinary spin-singlet
pairing when $N -Z > 10$.  The nucleus $^{132}_{60}$Nd is an example.
At the BdG minimum,
the only nonzero anomalous densities are the ones for $\alpha=1,3$ and they have
roughly equal amplitudes.  To see how the energy varies as the condensate
is changed, we constrain $\kappa_\alpha$ away from the values at the minimum
and examine the energy surface.  For a two-dimensional plot
we take the $x$ variable to be the amplitude of 
the neutron-neutron condensate $\kappa_1$ and the $y$ variable 
the amplitude of the neutron-proton condensate $\kappa_5$.
The amplitude of the proton-proton 
condensate $\kappa_3$ is taken to be the same as $\kappa_1$,
all other condensates are set to zero.  The resulting 
correlation energy for  $^{132}_{60}$Nd is shown as a contour plot in the left panel 
of Fig. \ref{fig:contour}. The peaks
near $(x,y) \approx (\pm 20,0)$ correspond to the unconstrained BdG minimum.  The
BdG energy does not depend on the phase of the condensate, so the
energy surface is symmetric under reflections in both axes.  The
pure uncorrelated ground state at the center of the graph (black dot)
defines the zero level for the condensation energy.  Note that the
contours are elongated in the vertical direction, indicating that
the energy surface is rather soft with respect to forming a spin-triplet
condensate.  

The middle panel of Fig. \ref{fig:contour} shows the energy 
surface for $^{132}_{66}$Dy on the $N=Z$ line.  Here the peaks are
at $(x,y) \approx (0, \pm 22)$, i.e. the condensate is spin-triplet.  Along the
$x$-axis the spin-singlet condensatation energy reaches a maximum
near $x=\pm20$,
but it is only a saddle point in the two-dimensional space.

We now ask how one condensate changes to the other
as $N-Z$ is varied.  One could imagine a sudden switch, corresponding
to a quantum phase transition, if the saddle point in the middle panel
of Fig. \ref{fig:contour} became
a peak that grew to become the global maximum.  This is not
what happens.  Instead, the peak shifts position, moving smoothly from
one axis to the other.  A typical case is shown in Fig. 3,
$N-Z=4$, i.e. the nucleus $^{132}_{64}$Gd.  The maxima are located at
$(x,y) \approx (\pm 11,\pm 20)$, i.e. the condensate has a mixed-spin character.  
We have examined the form of the pairing in the canonical basis 
and found that the relationship between paired orbitals $|i\rangle$ and
$|\bar i\rangle$ is $|\bar i\rangle  = \tau_z T | i \rangle$, where $T$
is the time reversal operator and $\tau_z$ is the Pauli isospin operator.
This particular form gives ordinary spin-singlet pairing in the absence of
 neutron-proton mixing in the orbital $|i\rangle$, but has a spin-triplet 
component when neutrons and
protons are mixed.

{\it Discussion.} Trial computations we have performed suggest that the phenomenon of mixed
pairing depends on the spin-orbit field in the nucleus. In the
absence of spin-orbit splitting, the singlet and triplet interactions
would be on an equal footing. There would be a sharp transition at 
$v_s=v_t$, the SU(4) symmetry point, with pure condensates of one type
or the other away from that point. Moreover, the smooth transition between spin-singlet and
spin-triplet pairing could not have been anticipated from results
such as those in Refs. \cite{sa97,ter98} on the behavior in $^{48}$Cr as a function of
angular momentum.  These authors found that the states of different character
do not mix strongly and the character of the yrast state just depends
on which is lower in energy.  Also, in Ref. \cite{sa01} the authors report
that there is a phase transition as a function of $N-Z$, with the
change of character taking place suddenly. However, a small degree
of mixing was found in Ref. \cite{go01} for the
high spin states in $^{80}$Zr.  It is important to note that the spin-triplet
condensates considered in these references are in channels different 
from the $\alpha=5$ one (see Table I) which we use.
In fact, we find a smooth transition to spin-singlet 
pairing only when starting from that channel, i.e. spin-triplet with $z$-projection 
$S_z=0$.

\begin{figure}[t]
\vspace{0.5cm}
\begin{center}
\includegraphics[width=0.46\textwidth]{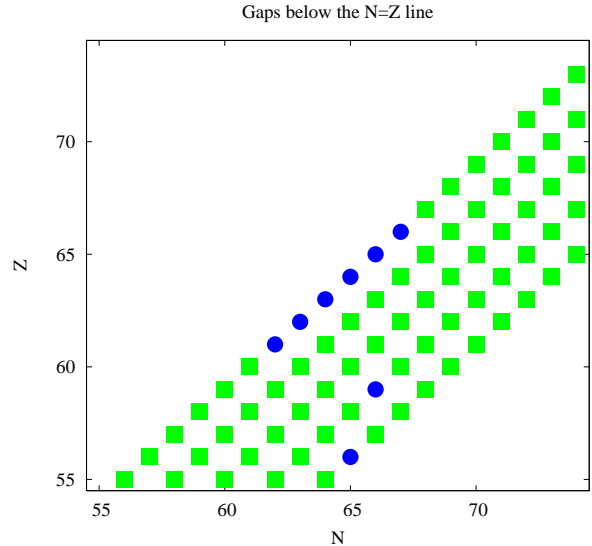}
\caption{(color online) 
Small and large pairing gaps (Eq. \ref{eq:gaps}) for nuclei near the spin-triplet pairing
region.  Blue circles:  $\Delta_o^{(3)}(n) < 0.4 $ MeV; green squares: $\Delta_o^{(3)}(n) >
0.4 $ MeV.  }
\label{fig:gaps}
\end{center}
\end{figure}

We now briefly mention some of the possible physical consequences of the mixed 
pairing phase.   One potential consequence
of spin-triplet pairing might be a reduced pairing gap in the odd-even
mass differences.  In particular, it was found in Ref.
\cite{be10} that some 
quasiparticle energies are close to zero, suggesting reduced pairing gaps.
To examine the pairing gaps we calculate the second difference of the 
correlation energies by the formula
\be
\Delta_o^{(3)}(n) = E(n) - \half \left [ E(n-1) + E(n+1) \right ] . 
\label{eq:gaps}
\ee
Here $n$ is either the neutron or proton number, taken to be odd, with
the other nucleon species held fixed at some even number.  Typical
values of the gap are 0.7-0.9 MeV in normally paired nuclei.  These
drop to 0.25 for some of the spin-triplet cases as shown in Fig.
\ref{fig:gaps}.  One sees that all of these nuclei
are one unit off the $N=Z$ line.  Farther off the line, where the 
pairing becomes mixed,
the gaps increase to the larger value.  Thus,
we find some evidence that the gaps are affected, but the predicted
effect does not extend within the drip line.

On a different note, there may be spectral signatures of the triplet
pairing and the transition.  According to Frauendorf, the spectra of
odd-odd nuclei become similar to that of even-even nuclei in the limit
of strong spin-triplet pairing (\cite{fr01}, p. 499).
condensation.  Even in the mixed-spin regime, the softness of the
energy surface suggests that there should be low-lying excitations
associated with the spin degree of freedom.  The mean-field theory has to be
extended to deal with broken symmetries including particle number and
angular momentum before predictions can be made for spectroscopic
quantities. Other observables of interest are two-particle transfer
direct reaction cross sections, which in principle can be used to compare
correlation strengths in the two spin channels.  Here, again, the theory
needs to restore good particle number to distinguish between the nuclei
participating in the transfer reaction, and this remains for future work.
Finally, nuclear deformation can strongly modify pairing behavior, so it
is important to extend the calculations to the full Hartree-Fock-Bogoliubov
theory in which deformation effects are included.
In this Letter, we have focused on the qualitatively new phenomenon 
of mixed-spin pairing, predicted for nuclei that are experimentally accessible.

\begin{acknowledgments} 
{\it Acknowledgments.} We thank M.~M. Forbes, P. Ring, and L.~M. Robledo for useful discussions.
This work was supported by U.S. DOE Grants No. DE-FG02-97ER41014 and No. DE-FG02-00ER41132.
\end{acknowledgments} 
 


\begin{thebibliography}{99} 
\bibitem{po98} A.~Poves and G.~Martinez-Pinedo, Phys. Lett. B{\bf 430}
203 (1998).
\bibitem{sc10} G.F.~Bertsch, A.O.~Macchiavelli, and A.~Schwenk,
unpublished; see also S. Baroni, A.O.~Macchiavelli, and A.~Schwenk,
Phys. Rev. C{\bf 81} 064308 (2010).
\bibitem{be10} G.F.~Bertsch and Y.~Luo, Phys. Rev. C{\bf 81} 064320 (2010).
\bibitem{goo72} A.~L. Goodman, Nucl. Phys. A {\bf 186} 475 (1972).
\bibitem{goo99} A.~L. Goodman, Phys. Rev. C {\bf 60} 014311 (1999).
\bibitem{ma00a}A.O.~Macchiavelli, et al., Phys. Rev. C {\bf 61} 041303R
(2000).
\bibitem{ma00b}A.O.~Macchiavelli, et al., Phys. Lett. B {\bf 480} 1
(2000).
\bibitem{sa97} W. Satula and R. Wyss, Phys. Lett. B {\bf393} 1 (1997).
\bibitem{ter98} J. Terasaki, R. Wyss, and P.-H. Heenen, Phys. Lett. B {\bf 437} 1
(1998).
\bibitem{go01} A.~L. Goodman, Phys. Rev. C {\bf 63} 044325 (2001).
\bibitem{goo98} A.~L. Goodman, Phys. Rev. C {\bf 58} 3051 (1998).
\bibitem{RS} P. Ring and P. Schuck, ``The nuclear many-body
problem'', (Springer, New York, 1980).
\bibitem{ro11} Our numerical procedures are introduced in L.M.~Robledo and
G.F.~Bertsch, to be published (2011); see also J. L. Egido, J. Lessing, 
V. Martin, and L. M. Robledo, Nucl. Phys. A{\bf 594}, 70 (1995).
\bibitem{sa01} W. Satula and R. Wyss, Phys. Rev. Lett. {\bf 86} 4488 (2001).
\bibitem{fr01} S. Frauendorf, Rev. Mod. Phys. {\bf 73} 463 (2001).
\bibitem{mo95} P. Moller, et al., At. Data Nucl. Data Tables {\bf 59} 185 (1995).
\end{thebibliography}
\end{document}